\documentclass[english,letter,12pt,fleqn]{article}

\usepackage{ulem} % added by Anne to make corrections
\usepackage{babel}
\usepackage{amsmath,amssymb,amsfonts,hhline}
\usepackage{graphicx}
\usepackage[hmargin=2.5cm,vmargin=2cm]{geometry}
\usepackage{natbib}

\usepackage{amsthm}
\usepackage{mathrsfs}
\usepackage{dpfloat}

\allowdisplaybreaks[1]
\newtheorem{theorem}{Theorem}[section]

\newtheorem{lemma}{Lemma}[section]

\newtheorem{definition}{Definition}[section]

\newtheorem{remark}{Remark}[section]

\providecommand{\diag}{\mathrm{diag}}
\providecommand{\off}{\mathrm{off}}
\providecommand{\argmin}{\mathrm{argmin}}
\providecommand{\arginf}{\mathrm{arginf}}

\title{\huge{\bf On asymptotics of ICA estimators and their performance indices}}

\author{Pauliina Ilmonen, Klaus Nordhausen, Hannu Oja, Esa Ollila \footnote{Pauliina Ilmonen is a Postdoc, Universit\'{e} libre de Bruxelles, Belgium, (email: Pauliina.Ilmonen@gmail.com). Klaus Nordhausen is a University Lecturer, University of Tampere, Finland (email: Klaus.Nordhausen@uta.fi). Hannu Oja is an Academy Professor, University of Tampere, Finland (email: Hannu.Oja@uta.fi).
Esa Ollila is a Academy Research Fellow, Aalto University School of Science and Technology, Finland (email: esa.ollila@aalto.fi)} \\ }

\date{Dec 12, 2012}
\begin{document}
\maketitle

\begin{abstract}

 Independent component analysis (ICA) has become a popular multivariate analysis and signal processing technique with diverse applications.
 This paper is targeted at discussing theoretical large sample properties of ICA unmixing matrix functionals.
We provide a formal % (mathematical)
definition of unmixing matrix functional %$\Gamma(F)$.
and  consider  two popular  estimators in detail:  the family based on  two scatter matrices with the
independence property  (e.g., FOBI estimator)  and the family of deflation-based fastICA
estimators. The limiting behavior of the corresponding estimates is
discussed and the asymptotic normality of the deflation-based fastICA estimate is proven under general assumptions.
Furthermore, properties of several performance indices commonly used for comparison of different unmixing matrix estimates are discussed and a new performance index is proposed.
The proposed index fullfills three desirable features which promote its use in practice and distinguish it from others. Namely,
the index possesses an easy interpretation, is fast to compute and its asymptotic properties can be inferred from asymptotics of the unmixing matrix
estimate. We illustrate the derived asymptotical results  and the use of the proposed index
with a small simulation study.\newline

\newline
\end{abstract}

\section{Introduction}
In the independent component (IC) model we assume that the
components of the observed $p$-variate random vector $x=(x_1,...,x_p)^T$ are
linear combinations of the components of a latent $p$-vector $z=(z_1,...,z_p)^T$
such that $z_1,...,z_p$  are mutually independent. Then
\begin{equation}\label{model}
     x=\Omega  z
\end{equation}
where $\Omega$ is a full-rank $p\times p$ mixing matrix. The model is semiparametric
as we do not make any assumptions on the marginal distributions of $z_1,...,z_p$. In order to
be able to identify a mixing matrix one has to assume that at most
one of the components $z_1,...,z_p$ is normally distributed, \cite{IN}. Still after this assumption,
the parameter matrix $\Omega$ is  not uniquely defined: Let $\mathcal C$ be the set
of $p\times p$ matrices with exactly one non-zero element in each
row and in each column. If $C\in\mathcal C$  then also $z^*=Cz$ has
independent components and the model can be rewritten as $x=\Omega^*
z^*$ where $\Omega^*=\Omega C^{-1}$.

There are several possible, but not always satisfactory,  solutions
to this identifiability problem. One then fixes $C$ by fixing either $z^*=Cz$ or
$\Omega^*=\Omega C^{-1}$ in some way.  First, the random vector $z^*=Cz$ can be
fixed by requiring, for example, that the components of $z^*$ satisfy
(i) $Var(z^*_i)=1$, $i=1,...,p$, (ii) $\beta_1(z^*_i)>0$,
$i=1,...,p$, and (iii) $\beta_2(z^*_1)>...> \beta_2(z^*_p)$ where
$\beta_1$ and $\beta_2$ are classical moment-based skewness and
kurtosis measures, respectively.
 The above idea was extended in \cite{Ilmonen:2010,NordhausenOllilaOja:2011b} by fixing the random vector $Cz$ using two different
location vectors and two different scatter matrices with the so
called independence property. In these approaches, some
indeterminacy still remains for random vectors with identical or
symmetrical marginal distributions, for example. Second, the transformation
matrix $\Omega^*=\Omega C^{-1}$ can be fixed by requiring, for example, that
$||\Omega C^{-1}-I_p||$ is minimized.
\cite{icaIP} used a unique representation $\Omega C^{-1}$ such that
all the diagonal elements of $\Omega C^{-1}$ are one.  In this paper
we accept the ambiguity in the model (\ref{model}), and try to define
our concepts and analysis tools so that they are independent of the
model specification, that is, of the specific choices of $z$ and $\Omega$.

In the independent component analysis (ICA) the aim is to find an
estimate for an unmixing matrix $\Gamma$ such that $\Gamma x$ has
independent components. Again, if $\Gamma$ is a mixing matrix then
so is $C\Gamma$ for all possible matrices $C\in \mathcal C$. Thus
$\Gamma=\Omega^{-1}$ in the model (\ref{model}) is just one possible
unmixing matrix and the ICA problem reduces to estimating an
unmixing matrix $\Omega^{-1}$ only up to the order, signs and scales
of the rows of $\Omega^{-1}$. In the signal processing and computer science
communities ICA procedures are usually seen as algorithms rather than estimates with their
statistical properties. The most popular algorithms, if formulated with random
variables, then often proceed as follows.
\begin{enumerate}
\item
In the model (\ref{model}), one can assume without loss of
generality that $Cov(z)=I_p$. Then, after whitening, we get the
random vector
\[
y=Cov(x)^{-1/2}(x-E(x))=V(z-E(z))
\]
with some orthogonal matrix $V$.
\item
Using $y$, find a orthogonal matrix $U=(u_1,...,u_p)^T$  with the
rows $u_i$, $i=1,...,p$, chosen to maximize (or minimize)
a criterion function, say $\sum_{i=1}^p |E[G(u_i^Ty)]|$. The
optimization may be conducted one by one  or simultaneously.  The
function $G$ (measure of non-gaussianity, negentropy,
kurtosis measure, log-likelihood function, etc.) is chosen so
that the solution is  $U=V^T$ up to possible sign changes and permutations of the rows.
\item
  The final ICA solution is then  $\Gamma=UCov(x)^{-1/2}$.
\end{enumerate}

The fastICA algorithms described  in \cite{HyvarinenOja:1997} for
example works in this way.  The rows of $U$ are then found either
one after another (deflation-based fastICA) or simultaneously
(symmetric fastICA). The sample versions are naturally obtained by
replacing the expectations by corresponding sample averages.
For detailed descriptions of the  fastICA
procedures and several other estimates and algorithms,
 see \cite{Cichocky2006} and \cite{IN}.
For other type of estimates, see \cite{ChenBickel:2005} and
\cite{ChenBickel:2006}.
% A nice discussion about the prewhitening step and under which conditions the existence of moments can then be neglected is found
%in \cite{ChenBickel:2005}. ICA algorithms following a different approach are especially described in \cite{ChenBickel:2006}.

Due to the vast amount of different ICA estimates and algorithms,
asymptotic as well as finite sample criteria are needed for their
comparisons. While results on asymptotic statistical properties
(convergence, asymptotic normality, etc.) are usually missing in the literature, several
finite-sample performance indices have been proposed  for the comparisons in simulation studies. Let
$\hat\Gamma$ be an unmixing matrix estimate based on the random
sample $X=(x_1,...,x_n)^T$ from the distribution in model
(\ref{model}). First, one can compare the ``true'' sources $z_i$
(which are of course known in the simulations) and the estimated
sources $\hat{z_i}= \hat{ \Gamma}x_i$, $i=1,...,n$. Second, one can
measure the closeness of the ``true'' unmixing matrix $\Omega^{-1}$
(used in the simulations) and the estimated unmixing matrix
$\hat{\Gamma}$. In both cases the problem is that $\hat{\Gamma}$ is
typically not an estimate for $\Omega^{-1}$. However, for any
reasonable estimate $\hat\Gamma$, either (i) there exists a $C\in \mathcal C$ such that
$\hat\Gamma$ is a consistent estimate of $\Omega C^{-1}$, or (ii) there exists a
(possibly unknown or unspecified) matrix $\hat C\in\mathcal C$ such that $\hat
C\hat\Gamma$ is a consistent estimate of $\Omega^{-1}$. Therefore,
for a good estimate, the gain matrix $\hat G=\hat\Gamma \Omega$
tends to be close to some matrix $C\in \mathcal C$. In this paper we
discuss performance indices  that are based on the use of $\hat
G=\hat\Gamma\Omega$. A new index is proposed that finds the shortest
distance (using Frobenius norm) between the identity matrix and the
set of matrices equivalent to the gain matrix $\hat{\Gamma} \Omega$.

We organize the paper as follows. First, in Section
\ref{ICfunctionals}, we give a formal (mathematical) definition of
the IC functional which is independent of the model formulation.
We consider two families  of  IC
functionals, (i) the family based on  two scatter matrices with
independence property, and (ii) the family of deflation-based
fastICA functionals. We review limiting behavior of the
corresponding estimates and we prove the asymptotic normality of the
deflation-based fastICA under certain general assumptions. Previous attempts to prove the asymptotic normality of the
deflation-based fastICA that have been presented in the literature contain severe faults.
In Section \ref{index} we  consider the use of the gain matrix in the comparison of different IC estimates.
Several approaches are discussed in detail.
 In Section
\ref{NewIndex} a new index for the comparison is introduced. The computation of the new index is shown to be
straightforward and easy. We also consider the limiting
behavior of the index as the sample size approaches infinity; the
asymptotic properties of the index are in a natural way determined
by the asymptotic properties of the estimate $\hat{\Gamma}$. The finite sample vs. asymptotic behavior
of the index for several different ICA estimates with known
asymptotics is illustrated in a small simulation study. Most proofs of the theorems
are placed in the Appendix.

\section{IC functionals}\label{ICfunctionals}
In this section we give a formal (mathematical) definition of an independent component (IC)
functional. The definition is independent of the model formulation,
that is, of the choice of $\Omega$ and $z$. As an example we
consider the family of IC functionals based on two scatter matrices
with independence property, and the family of deflation-based
fastICA functionals.

\subsection{Formal definition}
Let $\mathcal G$ be the set of all full-rank $p\times p$ matrices.
Then naturally all unmixing matrices $\Gamma\in \mathcal G$. Let $
P$ denote a permutation matrix (obtained from $I_p$ by permuting its
rows or columns), $J$ a sign-change matrix (a diagonal matrix with
diagonal elements $\pm 1$), $D$ a rescaling matrix (a diagonal
matrix with positive diagonal elements). For the definition of an IC
functional we need the subset
\begin{eqnarray*}
\mathcal C&=&\left\{ C \in\mathcal G: C = PJD \ \ \mbox{for some
$P$, $J$, and $D$} \right\}.
% \\&=&\left\{ C \in\mathcal G: C = PL \
%\ \mbox{for some $P$ and $L$} \right\}\\
%&=&\left\{ C \in\mathcal G: C = MPL \
%\ \mbox{for some $M,$ $P,$ and $L$} \right\}\\
%&=&\left\{ C \in\mathcal G: C = MP \
%\ \mbox{for some $M$ and $P$} \right\}.
\end{eqnarray*}
If $C\in \mathcal C$, each row and each column of $C$ has exactly one nonzero element.
Then $\mathcal C$ gives a group of affine
transformations (with respect to matrix multiplication)
as it satisfies  (i) if $C_1,C_2\in \mathcal C$ then $C_1C_2\in \mathcal C$, (ii)
$I_p\in \mathcal C$, (iii) if $C\in\mathcal C$ then there exists $C^{-1}\in \mathcal C$ such that
$CC^{-1}=C^{-1}C=I_p$. The group is not commutative (Abelian) as $C_1C_2=C_2C_1$ may not be true.

%$\bo\Gamma$ is an unmixing matrix if $\bo\Gamma \bo\Omega \sim \bo
%I_p$.
%\newline
We say that two matrices $\Gamma_1$ and $\Gamma_2$ in $\mathcal G$ are equivalent if
$\Gamma_1=C\Gamma_2$ for some $C\in\mathcal C$. We then write $\Gamma_1\sim \Gamma_2$ and
 give the following definition.
\begin{definition}\label{icdef}
Let $F_{x}$ denote the cdf of $x$ The functional $\Gamma(F_{x})\in
\mathcal G$ is an IC functional in the IC model (\ref{model}) if (i)
$\Gamma(F_{x}) \Omega\sim I_p$ and if (ii) it is affine equivariant
in the sense that $\Gamma(F_{Ax})=\Gamma(F_{x}) A^{-1}$ for all
nonsingular $p \times p$ matrices $A$.
\end{definition}

\begin{remark}\label{ICremark}
The first condition says that $\Gamma(F_x)$ and $\Omega^{-1}$ are equivalent
matrices and that there exists
$C=C(F_x,\Omega) \in\mathcal C$ such that the ``adjusted'' IC functional
$C \Gamma(F_x)=\Omega^{-1}$.
Note that, if the second condition (ii) is replaced by
a weaker condition (iii) $\Gamma(F_{Ax})\sim\Gamma(F_{x}) A^{-1}$ for
all nonsingular matrices $A$, then one can often find a new
functional
$$\Gamma^*(F_{x})=C(F_{x})\Gamma(F_{x})$$
with $C(F_{x})\in \mathcal C$  satisfying condition (ii). If the fourth moments exist, functional
$C=C(F_{x})$ may be defined by requiring, for example, that
$Var((\Gamma^*x)_i)=1$, $i=1,...,p$,  $\beta_1((\Gamma^*x)_i)>0$,
$i=1,...,p$, and  $\beta_2((\Gamma^*x)_1)>...> \beta_2((\Gamma^*x)_p)$ where
$\beta_1$ and $\beta_2$ are classical moment-based skewness and
kurtosis measures, respectively. Then
$\Gamma^*(F_{Ax})=\Gamma^*(F_{x}) A^{-1}$ for all nonsingular $p
\times p$ matrices $A$. Other criteria for constructing $C(F_x)$ can be easily found.
\end{remark}

\begin{remark}\label{ICremark2}
In practice, the IC functional is  often  seen rather as a set of vectors $\{\gamma_1,...,\gamma_p\}$  than as
a matrix $\Gamma=(\gamma_1,...,\gamma_p)^T$. If $P_i=||\gamma_i||^{-2} \gamma_i\gamma_i^T$ is the projection matrix to the subspace
spanned by $\gamma_i$, $i=1,...,p$, then the functional can also be defined as a set of projection matrices $\{P_1,...,P_p\}$.
\end{remark}

Note that, for an IC functional $\Gamma$ in model (\ref{model})
$\Gamma(F_{x}) \Omega C \sim I_p$ for all $C\in C$. Therefore the
definition of the IC functional does not depend on the specific formulation
of the model (the choices of $\Omega$ and $z$). Also, $\Gamma(F_x)x=\Gamma(F_z)z$ where
$\Gamma(F_z)$ is in $\mathcal C$. If we choose $z^*=\Gamma(F_{z})z$
and $\Omega^*=\Omega\Gamma(F_{z})^{-1}$, then
$\Gamma(F_x)\Omega^*=I_p$. This formulation of the model is then
most natural (canonical) for functional $\Gamma(F_{x})$.

%%%%%%%%%%%%%%%%%%%%%%%%%%%%%%%%%%%%%%%%%%%%%%%%%%%%%%%%%%%%%%%%%%
%%%%%%%%%%%%%%%%%%%%%%%%%%%%%%%%%%%%%%%%%%%%%%%%%%%%%%%%%%%%%%%%%

%\section{Two families of IC functionals}

\subsection{Functionals based on two scatter matrices}\label{ICS} A
{\it scatter functional} $S(F_{x})$ is  a $p\times p$ -matrix-valued
functional which is positive definite and affine equivariant in the
sense that
$$S(F_{Ax+b})=AS(F_{x})A^T$$
for all nonsingular $p\times p$ matrices $A$ and for all
$p$-vectors $b$. A scatter functional $S$ is said to possess the
{\it independence property} if $S(F_{x})$ is a diagonal
matrix for all $x$ with independent components. Naturally, the
usual covariance matrix $$S_1(F_{x})=E\left((x-E(
x))(x-E(x))^T \right)$$ is a scatter matrix with the
independence property. Another scatter matrix with the independence
property is  the matrix  based on fourth moments, namely,
\[
S_2(F_{x})=\frac 1{p+2} E\left((x-E(x))(x-E(x))^TS_1(F_{x})^{-1}
(x-E(x))(x-E(x))^T\right) .\] For any scatter matrix $S(F_{ x})$,
its symmetrized version $$S_{sym}(F_{x})=S(F_{ x_1-x_2}),$$ where
$x_1$ and $x_2$ are independent copies of $x$, has the independence
property, \cite{Oja:2006, Tyler:2009}. For symmetrized M-estimators
and S-estimators, see \cite{Roelandt:2009, Sirkia:2007}.
%\newline

The IC functional $\Gamma(F_{x})$ based on the scatter matrix
functionals  $S_1(F_{x})$ and $S_2(F_{x})$ is
defined as a solution of the equations
\[
\Gamma S_1 \Gamma^{T}=I_p \ \ \mbox{and} \ \
\Gamma S_2 \Gamma^{T}=\Lambda \] where
$\Lambda=\Lambda(F_{x})$ is a diagonal matrix with
diagonal elements $\lambda_1\ge ... \ge\lambda_p>0$.   One of the first solutions for the ICA problem, the FOBI
functional, \cite{FOBI},  is obtained if the scatter functionals $
S_1$ and $S_2$ are the scatter matrices based on the second and
fourth moments, respectively. The use of two scatter matrices in ICA
has been studied in \cite{Nordhausen:2008, Oja:2006} (real data) and
in \cite{Ollila:2008, Ilmonen:2012} (complex data).
%\newline

Assume now (wlog) that $\Omega=I_p$ and that $S_1(F_{ x})=I_p$ and
$S_2(F_{x})=\Lambda$ where $\lambda_1> ... >\lambda_p>0$. Assume
also that both $S_1$ and $S_2$ have the independence property. Write
 $\hat{S_1}=S_1(F_n)$ and
$\hat{S_2}=S_2(F_n)$ (values of the functionals at the empirical cdf
$F_n$).  We then have the following result, \cite{Ilmonen:2010}.

\begin{theorem}
Assume that
\[
\sqrt{n}(\hat{S}_1-I_p)=O_p(1)  \ \ \mbox{and}\ \
\sqrt{n}(\hat {S}_2-\Lambda)=O_p(1),
\]
with $\lambda_1>...>\lambda_p>0$, and the estimates $\hat{\Gamma}$
and $\hat{\Lambda}$ are given by
\[
\hat {\Gamma} \hat {S}_1 \hat {\Gamma}^{T}=I_p \ \
\mbox{and}\ \ \hat {\Gamma} \hat {S}_2 \hat {\Gamma}^{T}=
\hat{\Lambda}.
\]
Then, there exists a sequence of estimators such that $\hat\Gamma\to_P I_p$,
\begin{eqnarray*}
\sqrt{n} \mbox{diag}(\hat\Gamma-I_p)&=&-\frac 12 \sqrt{n}
\mbox{diag}(\hat S_1-I_p)+o_p(1) \ \ \mbox{and}\\
 \sqrt{n}\ \off(\hat{\Gamma}-I_p)
&=& \sqrt{n}\ H\odot\left((\hat {S}_2-\Lambda)- (\hat {S}_1-I_p)
\Lambda\right) +o_p(1),
\end{eqnarray*}
where $H$ is a $p\times p$ matrix with elements
\[
H_{ij}=0,\ \mbox {if}\ i=j,\ \ \mbox{and} \ \
H_{ij}=(\lambda_i-\lambda_j)^{-1},\ \mbox {if}\ i\ne j.
\]
\end{theorem}

Above  $\off (\Gamma)=\Gamma-\diag(\Gamma),$ where $\diag(\Gamma)$ is
a diagonal matrix with the same diagonal elements as $\Gamma$, and
$\odot$ denotes the Hadamard (entrywise) product.
\cite{Ilmonen:2010} considered  the limiting distribution of the
FOBI estimate (with limiting covariance matrix) in more detail. It
is interesting to note that the asymptotic behavior of the diagonal
elements of $\hat\Gamma$ does not depend on $\hat S_2$ at all.
%\newline

%To find an IC functional by jointly diagonalizing $k>2$ scatter matrices with the indepdence property
%has recently been suggested in \cite{NordhausenGutchOjaTheis:2011}.
Approaches such as JADE, \cite{JADE}, or the matrix-pencil
approach, \cite{yeredor:2009}, (approximately) jointly diagonalize  two
or more data matrices (not necessarily scatter matrices). The asymptotic properties of these
estimates are typically however still unknown.

%%%%%%%%%%%%%%%%%%%%%%%%%%%%%%%%%%%%%%

\subsection{Deflation-based FastICA functionals}\label{fastica}

Our second example on families of IC functionals is given by the
deflation-based fastICA algorithm. FastICA is one of the  most
popular and widespread ICA algorithms. Detailed examination of
fastICA functionals are provided for example in
\cite{HyvarinenOja:1997} and \cite{Ollila:2010}. In
\cite{Ollila:2010}, the asymptotic covariance structure of the row
vectors of deflation-based fastICA estimate $\hat\Gamma$ is given in
closed form.  No rigorous proof of the asymptotic normality of the
fastICA estimate has  been presented in the literature so far; see
for example \cite{ShimizuEtAl:2006}. In this section we discuss the
conditions needed  for  the asymptotic normality of the
deflation-based fastICA estimate.

Assume that $x=\Omega z$ as in model (\ref{model}) with finite first
and second moments $E(x)=\mu$ and $Cov(x)=\Sigma$. In this approach
the first row of $\Gamma$ is obtained when a criterion function
$|E(G(\gamma^T (x-\mu)))|$ is maximized under the constraint
$\gamma^T\Sigma \gamma=1$. If we wish to find more than one source
then, after finding $\gamma_1,...,\gamma_{k-1}$, the $k$th source
maximizes $|E(G(\gamma^T (x-\mu)))|$ under the constraint
\[
\gamma_k^T\Sigma \gamma_k=1 \ \ \mbox{and}\ \ \gamma_j^T\Sigma
\gamma_k=0,\ \ j=1,...,k-1.
\]

If $G$ satisfies the condition
\[
|E(G(\alpha_1z_1+\alpha_2z_2))|\le \max(|E(G( z_1))|,|E(G(z_2))|)
\]
for all independent $z_1$ and $z_2$ such that $E(z_1)=E(z_2)=0$ and
$E(z_1^2)=E(z_2^2)=1$ and for all $\alpha_1$ and $\alpha_2$ such
that $\alpha_1^2+\alpha_2^2=1$, then the independent components are
found using the above strategy. It is easy to check that the
condition  is
true for the classical kurtosis measure $G(z)=z^4-3$, for example %Then the components are found in the
%kurtosis order.
\citep{BUG}.

Write $T(F)$ for the mean vector
(functional) and $S(F)$ for the covariance matrix (functional). The
$k$th fastICA functional $\gamma_k(F)$ then optimizes the Lagrangian
function
\[
|E[G(\gamma_k^T (x-T(F)))]|-\frac {\lambda_{kk}} 2(\gamma_k^T S(F)
\gamma_k-1)- \sum_{j=1}^{k-1} \lambda_{jk} \gamma_j^T S(F) \gamma_k,
\]
where $\lambda_{1k},...,\lambda_{kk}$ are the Lagrangian
multipliers. If $g=G'$ then one can easily show that (under general
assumptions) the functional $\Gamma=(\gamma_1,...,\gamma_p)^T$
satisfies the $p$ estimating equations
\[
E[g(\gamma_k^T(x-T(F))(x-T(F)))]  = S(F) \sum_{j=1}^k
\gamma_j\gamma_j^T  E[g(\gamma_k^T(x-T(F))(x-T(F)))],
\]
$k=1,...,p$. If $z=\Gamma x$ has independent components then
$\Gamma$ solves the above estimating equations. Note, however, that
the estimating equations do not fix the order of sources
$\gamma_1,...,\gamma_p$ anymore.

\begin{remark}
As mentioned before, the ICA
procedures are often seen as algorithms rather than estimates with
 statistical properties. The popular choices of $g$ for
practical calculations are {\it pow3}: $g(z)=z^3$, {\it tanh}:
$g(z)=tanh(z)$, and {\it gauss}: $g(z)=ze^{-z^2/2}$, for example. If
$E(x)=0$ then the fastICA algorithm for $\gamma_k$ uses the
iteration steps
\begin{enumerate}
\item $\gamma_k \leftarrow
\Sigma^{-1}E[g(\gamma_k^Tx)x]-E[g'(\gamma_k^Tx)]\gamma_k$
\item $\gamma_k\leftarrow \gamma_k-\sum_{j=1}^k
(\gamma_k^T\Sigma\gamma_j)\gamma_j$
\item $\gamma_k\leftarrow \gamma_k/\sqrt{\gamma_k^T\Sigma\gamma_k}$
\end{enumerate}
The sample version is  naturally obtained if the expected values are
replaced by the averages in the above formula. It is important to
note that it is not known in which order the components are found in
the above algorithm. The order depends strongly on the initial value
in the iteration.
\end{remark}

We next consider the limiting behavior of the sample statistic $\hat\Gamma$
based on a random sample $x_1,...,x_n$. %The results are only sketched here.
We assume that $E(x_i)=0$ and $Cov(x_i)=I_p$ and that the true value
is $\Gamma=I_p=(e_1,...,e_k)^T$. Write $\bar x$ and $\hat S$ for the
sample mean vector and sample covariance matrix, respectively. If
the fourth moments exist then $\sqrt{n}\mbox{vec}(\bar x, \hat
S-I_p)$ have a limiting multivariate normal distribution (CLT).
Write $\hat\Gamma=(\hat{\gamma}_1,...,\hat{\gamma}_p)^T$ for the
fastICA estimate of $\Gamma$. Write also
$$
\mu_k=E[g(e_k^Tx_i)], \ \ \lambda_k=E[g(e_k^Tx_i)e_k^Tx_i]$$
and
$$\tau_k=E[g'(e_k^Tx_i)e_k^T
x_i], \  \  \delta_k=E[g'(e_k^Tx_i)],
$$
$k=1,...,p$. We need later the assumption that $\lambda_k\ne
\delta_k$, $k=1,...,p-1$. (If $g(z)=z^3$, for example, this
assumption rules out the normal distribution.) For sample
statistics
\[
T_k=\frac 1n \sum_{i=1}^n (g(e_k^Tx_i)-\mu_k)x_i \ \ \mbox{and}\ \ \hat
T_k=\frac 1n \sum_{i=1}^n g(\hat{\gamma}_k^T(x_i-\bar x))(x_i-\bar x)
\]
we need the assumption that, using the Taylor expansion,
\begin{equation}\label{FastICAass}
\sqrt{n}(\hat T_k-\lambda_k e_k)= \sqrt{n}T_k-\tau_k e_k e_k^T\sqrt{n}\bar x+\Delta_k \sqrt{n}(\hat{\gamma}_k-e_k)+o_P(1)
\end{equation}
where $\Delta_k=E[g'(e_k^T x_i)x_ix_i^T ]$, $k=1,...,p$. Again, if
$g(z)=z^3$ and the sixth moments exist, then (\ref{FastICAass}) is
true and  $\sqrt{n}(\hat T_k-\lambda_k e_k)$ has a limiting
multinormal distribution. The estimating equations for the fastICA
solution $\hat\Gamma=(\hat{\gamma}_1,...,\hat{\gamma}_p)^T$ are then given by
\begin{equation}\label{esteq}
\hat T_k=\hat S [\hat{\gamma}_1\hat{\gamma}_1^T+...+\hat{\gamma}_k\hat{\gamma}_k^T]\hat T_k,\ \ \ k=1,...,p.
\end{equation}
If (\ref{FastICAass}) is true and $U_k=\sum_{j=1}^k e_je_j^T$ then
\begin{eqnarray*}
(I_p-U_k) \sqrt{n}(\hat T_k-\lambda_k e_k)= \lambda_k [
\sqrt{n}(\hat S-I_p)e_k &+& \sum_{j=1}^k e_je_k^T
\sqrt{n}(\hat{\gamma}_j-e_j)\\
&+& \sqrt{n} (\hat{\gamma}_k-e_k)]+o_P(1)
\end{eqnarray*}
and we get the following result.

\begin{theorem}\label{fastICAas}
Let $x_1,...,x_n$ be a random sample from the model (\ref{model})
with $\Omega=I_p$,  $E(x_i)=0$, and $Cov(x_i)=I_p$.  Let
$\hat\Gamma=(\hat{\gamma}_1,...,\hat{\gamma}_p)^T$ be the solution for
estimating equations in (\ref{esteq}), and the estimate satisfies
$\hat\Gamma\to_P I_p$. Then, under the general assumptions given
above,
\begin{eqnarray*}
% \nonumber to remove numbering (before each equation)
  \sqrt{n} \hat{\gamma}_{kl} &=& \frac 1 {\lambda_k-\delta_k} \left[ e_l^T \sqrt{n} T_k-\lambda_k \sqrt{n} \hat S_{kl} \right]+o_P(1), \ \ \ \ \mbox{for
  $l>k$} \\
\sqrt{n} (\hat{\gamma}_{kk}-1) &=& -\frac 12  \sqrt{n} (\hat S_{kk}-1)+o_P(1),
\ \ \ \ \mbox{and}
 \\
  \sqrt{n} \hat{\gamma}_{kl} &=& \sqrt{n} \hat{\gamma}_{lk}- \sqrt{n} \hat S_{kl}+o_P(1)
  \ \ \ \ \mbox{ for $l<k$}
\end{eqnarray*}
\end{theorem}

\begin{remark}
Theorem \ref{fastICAas} implies that, if  $\sqrt{n} (T_k-\lambda_k
e_k)$, $k=1,...,p$,  and $\sqrt{n}\mbox{vec}(\hat S-I_p)$ have a
joint limiting multivariate normal distribution then also the limiting
distribution of  $\sqrt{n}\mbox{vec}(\hat \Gamma-I_p)$ is
multivariate normal. Interestingly enough, the limiting distribution
of the estimated sources $\hat{\gamma}_1,...,\hat{\gamma}_p$ depends
on the order in which they are found. The limiting behavior of the
diagonal elements of $\hat\Gamma$ does not depend on the choice of
the function $g(z)$. The initial value for $\hat\Gamma$ in the
fastICA algorithm fixes the asymptotic order of the sources. %The
%source estimates can be robustified if the covariance matrix is
%replaced by a robust scatter matrix with the independence property.
%Note that the independence property of the used scatter functional
%is crucial. Any robust scatter functional can not be used, as was
%shown in \cite{Bry}.
For more details, see \cite{NordhausenIlmonenMandalOjaOllila:2011}.
\end{remark}

%We will provide more detailed examination of the asymptotic normality of the deflation-based fastICA, its covariance structure and covariance structure's dependence on the order of the found sources in an upcoming paper.

%%%%%%%%%%%%%%%%%%%%%%%%%%%%%%%%%%%%%%%%%%%%%%%%%%%%%%%%%%%%%%%
%%%%%%%%%%%%%%%%%%%%%%%%%%%%%%%%%%%%%%%%%%%%%%%%%%%%%%%%%%%%%

\section{On Performance Indices}\label{index}

Let $X=(x_1,...,x_n)^T$ be a random sample from the model
(\ref{model}) with some  choice of $\Omega$ and $z$. An estimate of the population quantity
$\Gamma(F_x)$ is obtained if the functional is applied to the sample cdf $F_n$. We then
write $\hat\Gamma$ or $\Gamma(F_n)$ or $\Gamma(X)$. The {\it gain matrix } $\hat G=\hat \Gamma \Omega$
is then generally used to compare the performances of different estimates.
For any reasonable estimate, $\hat G\to_P C$ for some $C=C(F_z)\in\mathcal C$.
How can one then compare matrices $\hat G$ converging to a different population value  $C$ that
depend on functional  $\Gamma$ and the specific choice of $\Omega$ and $z$ in the model (\ref{model})?

\subsection{Canonical parametrization}

For a comparison of different estimates $\hat\Gamma$ choose,
separately for each IC functional $\Gamma$, the corresponding canonical
parametrization
\[
x= \Omega^* z^*=(\Omega \Gamma(F_z)^{-1}) (\Gamma(F_z) z).
\]
Note that  $\Omega \Gamma(F_z)^{-1}$ does not depend on the model
formulation (the original  choices of $\Omega$ and $z$) at all and
that $\Gamma(F_x)\Omega \Gamma(F_z)^{-1}=I_p$. A correctly adjusted
gain matrix
\[
\hat G=\hat\Gamma\Gamma(F_x)^{-1}=\hat\Gamma\Omega \Gamma(F_z)^{-1}
\]
can then be used for a fair comparison of different estimates
$\hat\Gamma$ as in the model (\ref{model}) $\hat G\to_P I_p$ for all $\Gamma$ .
A natural performance index can then be defined as $D^2(\hat
G)$ where
\[
D^2(G)=||G-I_p||^2.
\]
If $\sqrt{n}vec(\hat G-I_p)\to_d N_{p^2}(0,\Sigma_\Gamma)$ (as is true
with the estimates in Sections  \ref{ICS} and \ref{fastica}) then
we get the following result.

\begin{theorem}\label{main-canonical}
Assume that, for the correctly adjusted gain matrix
\[
\hat G=\hat\Gamma\Omega \Gamma(F_z)^{-1},
\]
it holds  that
$\sqrt{n} \ \mathrm{vec} (\hat G-I_p)\to_d N_{p^2}(0,
\Sigma_\Gamma) $. Then the limiting distribution of
$
n\cdot D^2(\hat G)
$
 is that of
$\sum_{i=1}^k \delta_i\chi^2_i$ where
$\chi^2_1,....,\chi^2_k$ are independent chi squared variables with
one degree of freedom, and $\delta_1,...,\delta_k$ are the $k$
nonzero eigenvalues (including all algebraic multiplicities) of $\Sigma_\Gamma$.
\end{theorem}

\subsection{Adjusted functional}

It is often hoped that the independent components in $\Gamma(F_x)x$ are
standardized in a similar way and/or given in a certain order.
To formalize this step, we then need the following auxiliary functional
to standardize (rescale and reorder) the components.

\begin{definition}
Let $F_{x}$ denote the cdf of $x$ The functional $C(F_{x})\in
\mathcal C$ is a {\it standardizing functional} if it satisfies
\[
C(F_{Ax})=C(F_{x})A^{-1},\ \ \mbox{for all $A\in \mathcal C$}
\]
\end{definition}

\begin{remark}\label{ICremark3}
%A functional $C(F_x)$ is called an invariant coordinate (IC) functional if
%$C(F_{Ax})=C(F_{Ax})A^{-1}$, for all $A\in \mathcal G$, see \ref{Ilmonen:2011}. An IC functional is naturally a standardizing functional
%but this concept is too strong for our purposes here.
If the fourth moments exist, functional
$C=C(F_{x})$ may be defined by requiring, for example, that
$Var((Cx)_i)=1$, $i=1,...,p$,  $\beta_1((C x)_i)>0$,
$i=1,...,p$, and  $\beta_2((C x)_1)>...> \beta_2((Cx)_p)$ where
$\beta_1$ and $\beta_2$ are, as before, classical moment-based skewness and
kurtosis measures, respectively. Of course, the functional is not well defined
if the components have the same distribution. Note, however, that the corresponding sample statistic
is uniquely defined (with probability one). Other standardizing functionals can be easily found.
\end{remark}

\begin{definition}
Let $F_{x}$ denote the cdf of $x$, and $\Gamma(F_x)$ an IC functional.
Then the adjusted IC functional $\Gamma^*(F_x)$  based on $C(F_x)$ is
\[
\Gamma^*(F_x)=C(F_{\Gamma(F_x)x}) \Gamma(F_x).
\]
\end{definition}

Note that adjusted IC functionals are directly comparable as they
all estimate the same population quantity. The estimate is
\[
\hat\Gamma^*=C(X\Gamma(X)^T)\Gamma(X)
\]
and the gain matrix reduces to
\[
\hat G=\hat\Gamma^* \Gamma^*(F_x)^{-1}=C(X\Gamma(X)^T)\Gamma(X) \Omega C(F_z)^{-1}.
\]

The standardizing functional $C(F)$ is thus needed to fix the scales, the signs, and the order of the estimated independent
components. The  rescaling part $D(F)$ of the functional $C(F)$ is a diagonal matrix with positive diagonal elements, and it is often
determined by a scatter functional $S(F)$ so that
\[
D(F_x)=\left(\mbox{diag} (S(F_x))\right)^{-1/2}.
\]
The rescaled IC functional is then $\Gamma^*(F_x)=D(F_{\Gamma(F_x)x}) \Gamma(F_x)$ with the sample version
\[ \hat \Gamma^*=\hat D \hat \Gamma\ \ \mbox{ where $\hat D=\left(\mbox{diag}(\hat \Gamma \hat S \hat\Gamma^T)\right)^{-1/2}$ }.\]
We next consider the effect of the rescaling functional.

\begin{theorem}\label{AdjEst}
Assume (w.l.o.g.) that $\Omega=I_p$ and  $S(F_z)=I_p$.
Assume that $\sqrt{n}(\hat S-I_p)=O_P(1)$ and $\sqrt{n}(\hat\Gamma-D^{-1})=O_P(1)$ for some
diagonal matrix $D$ with positive diagonal elements.
Write $\hat \Gamma^*=\hat D \hat \Gamma$ where $\hat D=\left(\mbox{diag}(\hat \Gamma \hat S \hat\Gamma^T)\right)^{-1/2}$.
Then
\[
\sqrt{n}(\hat \Gamma^*-I_p)=-\frac{\sqrt{n}}2  \mbox{diag}(\hat S-I_p)+
\sqrt{n}\mbox{off}(D\hat\Gamma-I_p)+o_P(1).
\]
\end{theorem}

The gain matrix for the comparisons is thus
\[
\hat G=\hat \Gamma^* \Omega D(F_z)^{-1}
\]
with the limiting distribution given by Theorem \ref{AdjEst}.
As, for all the estimates $\hat\Gamma^*$, the limiting behavior of the diagonal elements of $\hat G$ is similar, one can use
$||\mbox{off}(\hat G)||^2$
in the comparisons.
If $\sqrt{n}vec(\hat G-I_p)\to_d N_{p^2}(0,\Sigma_{\Gamma^*})$  then
we get the following result.

\begin{theorem}\label{main-AdjEst}
Assume that, for the  gain matrix of the adjusted estimate
\[
\hat G=\hat \Gamma^* \Omega D(F_z)^{-1},
\]
it holds  that
$\sqrt{n} \ \mathrm{vec} (\hat G-I_p)\to_d N_{p^2}(0,
\Sigma_{\Gamma^*}) $. Then the limiting distribution of
$
n ||\mbox{off}(\hat G)||^2
$
 is that of
$\sum_{i=1}^k \delta_i\chi^2_i$ where
$\chi^2_1,....,\chi^2_k$ are independent chi squared variables with
one degree of freedom, and $\delta_1,...,\delta_k$ are the $k$
nonzero eigenvalues (including all algebraic multiplicities) of
\[
(I_{p^2}-D_{p,p})
\Sigma_{\Gamma^*} (I_{p^2}-D_{p,p}),
\]
with $D_{p,p}=\sum_{i=1}^p (e_i e_i^{T})\otimes (e_i
e_i^{T})$.
\end{theorem}

\subsection{Solution as a set $\{\hat\gamma_1,...,\hat\gamma_p\}$}

Note that the first two approaches above do not depend on how we fix $\Omega$ and $z$
in the model (\ref{model}). In these two approaches it is assumed, however,  that $\hat\Gamma\Omega$ is a root-$n$ consistent
estimate of some $C\in \mathcal C$. Among other things, this means that the order, signs, and scales of
the independent component functional are fixed in some way. In practice, the solution in the ICA problem is often seen rather as a set
$\{\hat\gamma_1,...,\hat\gamma_p\}$ than a matrix $\hat\Gamma=(\hat\gamma_1,...,\hat\gamma_p)^T$. The vectors $\hat\gamma_j$, $i=1,...,p$,
span corresponding univariate linear subspaces; thus the order, signs and lengths of $\hat\gamma_j$ are not interesting. Finally,
in the comparisons, one is usually only interested in the set of gain vectors $\{\hat g_1,...,\hat g_p\}$
where $\hat g_i=\Omega^T\hat \gamma_i$, $i=1,...,p$, not in the gain matrix $\hat G=(g_1,...,g_p)^T$ itself.

A common way to standardize the lengths of the rows of the gain matrix
is to transform $\hat G\to
\hat D \hat G$ where $\hat D$ is a diagonal matrix with
diagonal elements
\begin{equation}\label{stand2}
\hat D_{ii}=\left\{\max_j |\hat G_{ij}|\right\}^{-1},\ \ i=1,...,p.
\end{equation}
 We then have the following result.

\begin{theorem}\label{stand2_Th}
Assume that $\Omega=I_p$ and that
$\sqrt{n}(\hat G- D^{-1})=O_P(1)$ where $D$ is a diagonal matrix with positive diagonal elements. Let $\hat D$ be a diagonal matrix
given (\ref{stand2}). Then
\[
\sqrt{n}(\hat D\hat G -I_p)=
\sqrt{n}\mbox{off}\left(D\hat
G-I_p \right)+o_P(1) .
\]
\end{theorem}

The inference-to-signal (ISR) ratio and inter-channel inference
(ICI), \cite{douglas:2007}, uses this row-wise consideration and is
given by
\[
\sum_{i=1}^p \left( \sum_{j=1}^p \frac {\sum_{j=1}^p \hat
G_{ij}^2}{\max_j \hat G_{ij}^2}-1 \right).
\]
This index is invariant under permutations and sign changes of the
rows (and columns) of  $\hat G$, and it is also naturally invariant under
heterogeneous rescaling of the rows. It depends on the choice of
$\Omega$ but no adjustment of  $\hat\Gamma$ is needed.
Theorem \ref{stand2_Th} can be used to find asymptotical properties of this criterion.

 One of the most popular performance indices, the Amari index, \cite{amari_etal:1996},
 is defined as
\[
\frac 1p \left[ \sum_{i=1}^p \frac {\sum_{j=1}^p |\hat
G_{ij}|}{\max_j |\hat G_{ij}|} + \sum_{j=1}^p \frac {\sum_{i=1}^p
|\hat G_{ij}|}{\max_i |\hat G_{ij}|}
 \right]-2.
\]
 The index is invariant under permutations and sign changes of the
rows and columns of $\hat G$. However, heterogeneous rescaling of
the rows (or columns) on $\hat G$ changes its value. Therefore, the rows of $\hat\Gamma$ should be
rescaled in a suitable way and use $\hat G=\hat D \hat\Gamma
\Omega$. (A general practice in the signal processing community is that $\Omega$ and $z$ are chosen so
that $Cov(z)=I_p$ and that the sample covariance matrix of
$\hat\Gamma x_1,..., \hat\Gamma x_n$ is $I_p$ as well.) However, as
the index is based on the $L_1$ norm, its limiting distribution is
quite complicated. The intersymbol interference (ISI),
\cite{moreau_macchi:1994}, is similar to the Amari index in that it
is also based  on similar row-wise and column-wise considerations
and that similar adjusting is needed for $\hat\Gamma$.

\bigskip

\cite{ChenBickel:2006} for example use an invariant
criterion by computing the norm $||\hat{\Gamma}\Omega - I_p ||$,
after suitable rescaling, sign changing, and permutation of the rows
of $\hat\Gamma$ and columns of $\Omega$.

\section{A new index for the comparison}\label{NewIndex}

\subsection{Minimum distance index}

Let $A$ be a $p\times p$ matrix. The shortest squared distance
between the set $\{CA\ :\ C \in \mathcal C\}$ of
equivalent matrices (to $A$) and $I_p$ is given by
\[
 D^2(A)
=\frac 1{p-1} \inf_{C\in \mathcal C} \| C A-I_p\|^2
\]
where $\|\cdot\|$ is the matrix (Frobenius) norm.
\begin{remark}\label{remark}
%Let $A$ be a $p\times p$ matrix, $C_1 \in \mathcal C$ and let $C_m= \arginf_{C\in \mathcal C} \| CA-I_p\|^2.$ Now clearly
%$$\arginf_{C\in \mathcal C} \| CC_1A-I_p\|^2=C_m (C_1)^{-1}$$ and thus $$ D^2(A)= D^2(C_1A)$$ for all $C_1 \in \mathcal C.$
Note that $D^2(A)=D^2(CA)$ for all $C\in \mathcal C$.
\end{remark}
%It is
%straightforward to see that
\begin{theorem}\label{fourcond}
Let $A$ be any $p\times p$ matrix having at least one nonzero element in each row. The shortest squared distance $D^2(A)$ fulfils the following four conditions:
\begin{enumerate}
\item $1\ge D^2(A)\ge 0$,
\item $D^2(A)=0$ if and only if $A \sim I_p$,
\item $D^2(A)=1$ if and only if $A \sim 1_p a^T$ for
some $p$-vector $a$, and
\item the function $c\to D^2(I_p + c \ \off(A))$ is increasing in $c\in [0,1]$ for all matrices $A$ such that
$A_{ij}^2 \leq 1$, $i\ne j$.
%The operator $\off(\cdot)$ picks all
%off-diagonal values of a square matrix and is defined as $\off(\bo
%A) = \bo A - \diag(\bo A)$.
\end{enumerate}
\end{theorem}
Let $X=(x_1,...,x_n)^{T}$ be a random sample from a distribution
$F_{x}$ where $x$ obeys the IC model (\ref{model}) with unknown
mixing matrix $\Omega$. Let $\Gamma(F)$ be an IC functional. Then
clearly $D^2(\Gamma(F_{x})\Omega)=0$. If $F_n$ is the empirical
cumulative distribution function based on $X$ then
\[
\hat{\Gamma}= \hat{\Gamma}(X)=\Gamma(F_n)
\]
is the unmixing matrix estimate based on the functional
$\Gamma(F_{x})$.

The shortest distance between the identity matrix and the set of
matrices $\{C \hat{\Gamma}\Omega: C\in \mathcal C\}$
equivalent to the gain matrix $\hat{G}=\hat{\Gamma}\Omega$
is as given in the following definition.

\begin{definition} The minimum distance index for $\hat{\Gamma}$ is
\[
\hat D=
 D(\hat{\Gamma} \Omega)
=\frac 1{\sqrt{p-1}} \inf_{C\in \mathcal C} \| C
\hat{\Gamma}\Omega -I_p\|.
\]
\end{definition}

It follows directly from Theorem \ref{fourcond}, that $1\ge\hat D\ge 0$, and $\hat D=0$ if and only if
$\hat{\Gamma}\sim {\Omega}^{-1}$. The worst case with $\hat D=1$ is
obtained if all the row vectors of $\hat{\Gamma}\Omega$ point to the
same direction. Thus the value of the minimum distance index is easy to interpret. Note that $D(\hat\Gamma\Omega)=D(C\hat\Gamma\Omega)$
for all $C\in\mathcal C$. Also,
\[
D(\Gamma(XA^T)A\Omega)=D(\Gamma(X)\Omega).
\]
 Note also the nice
and natural local behavior described  in Theorem \ref{fourcond}, condition
4.

\cite{Theis:2004} proposed an  index called the
generalized crosstalking error which is  defined as the shortest
distance $||\Omega-\hat{\Gamma}^{-1}C||$ between the mixing matrix
$\Omega$ and its adjusted estimate $\hat{\Gamma}^{-1}C$, $C\in
\mathcal C$. %Recall that the generalized crosstalking error in
%\cite{Theis:2004} is defined as
The generalized crosstalking error is then defined as
\[
E(\Omega,\hat{\Gamma})=\inf_{C\in \mathcal C}\| \Omega
-\hat{\Gamma}^{-1} C \|
\]
where $\|\cdot\|$ denotes a matrix norm. Clearly,
$E(\Omega,\hat\Gamma)=E(\Omega,C\hat\Gamma)$ for all $C\in \mathcal
C$ but $E(A\Omega,\Gamma(XA^T))=E(\Omega,\Gamma(X))$ is not
necessarily true. If the Frobenius norm is used, the new index  may
be seen as a standardized version of the generalized crosstalking
error as
\[
\hat D=\inf_{C\in \mathcal C}\| C^{-1} \hat{
\Gamma}\left(\Omega -\hat{\Gamma}^{-1} C \right)\|.
\]
Note
that, unlike the minimum distance index,  the values of the Amari
index for $\hat\Gamma\Omega$ and $D\hat\Gamma\Omega$ (with a
diagonal matrix $D$) may  differ. The Amari index thus silently
assumes that the rows of $\hat\Gamma$ are prestandardized in a
specific way. The minimum distance index is compared to other
indices in more detail in \cite{NordhausenOllilaOja:2010}.

%This is an advantage when compared to the Amari index, since the
%Amari index is not affine invariant and therefore the values of it
%depend on the model formulation \cite{Nordhausen:2008}. Therefore
%there might be pitfalls when different algorithms are compared using
%the Amari index.

\subsection{Computation}\label{comp} At
first glance the index $\hat D= D(\hat{\Gamma}\Omega)$ seems
difficult to compute in practice as the minimization is  over all
choices  $C\in \mathcal C$. However, the minimization can be done
by two easy steps.
\begin{lemma}\label{trace}
Let $\mathcal P$ denote the set of all $p\times p$ permutation matrices. Let  $\hat{G} =
\hat{\Gamma}\Omega$, and let $\tilde{ G}_{ij}={\hat
G_{ij}^2}/{\sum_{k=1}^p \hat G_{ik}^2}$, $i,j=1,...,p$. Now the
minimum distance index can be written as
\[
\hat D= D(\hat{G}) = \frac 1{\sqrt{p-1}} \left(p - \max_{
P\in \mathcal P}\left( \mbox{tr}(P \tilde{G})\right) \right)^{1/2}.
\]
\end{lemma}
The maximization problem
\[ \max_{P}\left( \mbox{tr}(P
\tilde{G})\right)\] over all permutation matrices $P$ can be
expressed as a linear programming problem where the constraints are that all rows and all columns must add up to 1. In a personal
communication Ravi Varadhan  pointed out that it can be seen also as
a linear sum assignment problem (LSAP). That LSAP, which is a special case of linear programming, is equivalent to finding a minimizing permutation matrix as is
stated for example in \cite[Chapter 8.5]{DantzigThapa1997}. The cost matrix $\Delta$ of
the LSAP in this case is given by $\Delta_{ij}= \sum_{k=1}^p (I_{jk}
- \tilde{G}_{ik})^2$, $i,j,=1,...,p$, and many solvers exist for the computation.
We used the Hungarian
method (see e.g. \cite{Papadimitriou:1982})  to find
the maximizer $\hat {P}$, and in turn compute $\hat D$ itself.

The ease of computations is demonstrated  in Table \ref{Computation
times} where we give the computation time of thousand indices for
randomly generated $p\times p$ matrices in different dimensions. The
computations were performed on an Intel Core 2 Duo T9600, 2.80 GHz,
4GB Ram using MATLAB 7.10.0 on Windows 7.

\begin{table}
  \centering
  \begin{tabular}{c|cccccc}
    % after \\: \hline or \cline{col1-col2} \cline{col3-col4} ...
    $p$ & 3 & 5 & 10 & 25 & 50 & 100 \\ \hline
    Time & 0.19 & 0.29 & 0.64 & 3.13 & 12.62 & 57.54 \\
  \end{tabular}

  \caption{Computation time in seconds for 1000 indices for different dimensions $p$.}\label{Computation times}
\end{table}

An R-implementation of the index is available in the R-package JADE, \cite{NordhausenCardosoOjaOllila:2011}.

\subsection{Asymptotic behavior} \label{Asymp}
Let the model be written as
$x=\Omega z,$
where now $z$ is standardized such that $\Gamma(F_{z})=
I_p$. Then $\Gamma(F_{x})=\Omega^{-1}$, and  without any
loss of generality we can  assume that
$\Gamma(F_{x})=\Omega= I_p.$
We then have the following.

\begin{theorem}\label{main}
Assume that the model is fixed such that $\Gamma(F_{
x})=\Omega= I_p$ and that $
\sqrt{n} \ \mathrm{vec} (\hat{\Gamma}-I_p)\to_d N_{p^2}(0,
\Sigma) $. Then
\[
n\hat D^2=\frac n{p-1} \|  \off(\hat{\Gamma})\|^2 + o_P(1)
\]
and the limiting distribution of $n\hat D^2$ is that of
$(p-1)^{-1}\sum_{i=1}^k \delta_i\chi^2_i$ where
$\chi^2_1,....,\chi^2_k$ are independent chi squared variables with
one degree of freedom, and $\delta_1,...,\delta_k$ are the $k$
nonzero eigenvalues (including all algebraic multiplicities) of \[ASCOV(\sqrt{n} \
\mbox{vec}(\off(\hat{\mathrm\Gamma})))=(I_{p^2}-D_{p,p})
\Sigma (I_{p^2}-D_{p,p}),
\]
with $D_{p,p}=\sum_{i=1}^p (e_i e_i^{T})\otimes (e_i
e_i^{T})$.
\end{theorem}

Note that, for the theorem, we fix the model in a specific way (canonical formulation, $\Gamma(F_z)=I_p$) to find the limiting distribution. Then, for all choices of of $\Omega$ and $z$,
$$
n\hat D^2= \frac n{p-1} || \mbox{off}(\hat \Gamma \Gamma(F_z)^{-1})||^2
+o_P(1)
$$
where $\hat \Gamma$ is as in Theorem \ref{main}.
Note also that
the mean of the limiting distribution of
$n(p-1)(\hat D^2)$ is equal to $tr\left(ASCOV(\sqrt{n} \
\mbox{vec}(\off(\hat{\Gamma})))\right),$ which is a regular
global measure of the asymptotic accuracy of the estimate
$\hat{\Gamma}$ in a model where it is estimating the identity
matrix. Furthermore, to calculate this limiting value, it
is enough to know the asymptotic variances of  elements of
$\hat\Gamma$ only. Recall  that the variances of diagonal elements
are not used.

It is also important to note that similar asymptotical results for
the Amari index cannot be found  since (i) it is not invariant in
the sense that the values for $\hat\Gamma\Omega$ and
$D\hat\Gamma\Omega$ may differ, and (ii) it is based on the use of
$L_1$ norms.

\begin{remark} \label{other}
The new performance index presented in this paper is based on
$$\inf_{C\in \mathcal C} \| C
\hat{\Gamma}\Omega -I_p\|.$$ This formulation can be seen as a method that fixes the mixing matrix $\Omega$ and transforms $\hat{\Gamma}$ to
optimally adjusted $\hat{C}\hat{\Gamma}$. The index is not invariant
under the transformations $\Omega\to \Omega C^{-1}$. One could alternatively base the index on
$$\inf_{C\in \mathcal C} \|
\hat{\Gamma}\Omega C -I_p\|.$$ This alternative formulation can be seen as a method that fixes the unmixing matrix estimate $\hat{\Gamma}$ and transforms $\Omega$ to
optimally adjusted (random) $\Omega \hat{C}^{-1}$.  Asymptotical behavior of this index is similar to that of the minimum distance index $\hat D$ but it is not
invariant under transformations $\hat\Gamma\to C\hat\Gamma$. It seems more natural to us to fix $\Omega$ and $z$ and allow transformations to $\hat\Gamma$.
\end{remark}

\begin{remark}
Still another interesting possibility is to define the criterion index as
$$\inf_{C_1, C_2\in \mathcal C} \|
C_1\hat{\Gamma}\Omega C_2^{-1} -I_p\|.$$
This index is naturally invariant under both $\hat\Gamma\to C_1\hat\Gamma$ and $\Omega\to \Omega C_2^{-1}$ and is fully model independent.
Unfortunately, it does not seem to work in practice. In the bivariate case, for example, it is easy to see that, for all choices of $g_{11}\ne 0$, $g_{22}\ne 0$ and $g_{21}$, the gain matrices
\[
\hat{\Gamma}\Omega=
\left( \begin{array}{cc}
g_{11}& 0\\
g_{21} & g_{22}
\end{array} \right)
\]
all give the optimal index value zero.
\end{remark}

\subsection{A simulation study} \label{simu}
The finite-sample behavior of the new index $\hat D$  is now
considered for three estimates, namely, (i) the FOBI estimate, and
(ii) the deflation based fastICA  with $g(z)=z^3$  (pow3), and (iii)
the deflation based fastICA  with  $g(z)=tanh(z)$ (tanh). The
asymptotic normality of the FOBI estimate is proven in
\cite{Ilmonen:2010}. See  \cite{Ilmonen:2010} also for the limiting
covariance matrix of the FOBI estimate.  The asymptotic covariance
matrix of the deflation based fastICA estimate is given in
\cite{Ollila:2010}. Asymptotic normality was proven in this paper.
If the parametric marginal distibutions were known, it is possible
to find the maximum likelihood estimate (MLE) of the unmixing
matrix;  \cite{OllilaKimKoivunen:2008} found its limiting covariance
matrix. As a general reference value we can then compute the
Cramer-Rao type lower bound for $tr\left(ASCOV(\sqrt{n} \
\mbox{vec}(\off(\hat{\Gamma})))\right)$.

The simulation setup consists of three ($p=3$) independent
components with Laplace, logistic and $beta(3,3)$ distributions.
They were all standardized to have expected value 0 and variance 1.
In the simulations, the  mixing matrix $\Omega$ was the identity
matrix $I_3$. The sample sizes  were $n=5000$, $10000$, $25000$,
$50000$, $75000$, $100000$ with $10000$ repetitions, and for each
repetition the value of $\hat D$ was computed for all the estimates.
As shown   for the  fastICA estimates in Section \ref{fastica}, the
limiting distribution of the estimated sources
$\hat{\gamma}_1,...,\hat{\gamma}_p$ depends on the order in which
the algorithm finds them. In practice, the order can be controlled
with the initial value of the algorithm. Using the identity matrix
as initial value, for example,  finds the sources in the order they
are given above, and a permuted identity matrix as a starting value
finds the sources  in a similarly permuted order. To illustrate this
property in our simulations, we extracted the sources in two
different orders, (a) $beta(3,3)$, logistic and Laplace, and (b)
Laplace, logistic and $beta(3,3)$. The estimates are then denoted by
pow3(a), pow3(b), tanh(a), and tanh(b), respectively.
%(For the asymptotics it makes no
%difference, if the starting point of the algorithm is a permuted
%identity matrix or if the original sources are simulated in permuted
%order and the starting point is the identity matrix.)
%To illustrate that, fastICA estimates were calculated using first sources in order  and identity matrix as a starting point of the algorithm, in which case the sources were %found in the original order. We denote the estimates by pow3 a and tanh a.  Then fastICA estimates were calculated using using permuted the order $beta(3,3)$, logistic and %Laplace and again identity matrix as a starting point of the algorithm. We denote the estimates by pow3 b and tanh b.

Using the results in \cite{Ilmonen:2010}, one can calculate the
limiting variances of the components of
$\sqrt{n}(\hat{\Gamma}_{FOBI}-I)$. As a matrix form, the variances
then are
$$V_{FOBI}=\left( \begin{array}{ccc}
1.25& 26.71 & 5.07\\
24.38 & 0.80&8.78\\
4.43& 8.51&0.33
\end{array} \right)$$
where $(V_{FOBI})_{ij}$ is the limiting variance of
$\sqrt{n}(\hat{\Gamma}_{FOBI}-I)_{ij}$, $i,j=1,2,3$. Then
\begin{eqnarray*}
% \nonumber to remove numbering (before each equation)
  tr\left(ASCOV(\sqrt{n}\
\mbox{vec}(\off(\hat{\Gamma}_{FOBI})))\right) &=& 24.38+4.43+26.71\\ &+& 8.51+5.07+8.78 \\
   &=&77.88.
\end{eqnarray*}
Similarly, using results in   \cite{Ollila:2010},
%The asymptotic
%variance matrices of $\sqrt{n}(\hat{\Gamma}_{pow3 a}-I)$ and
%$\sqrt{n}(\hat{\Gamma}_{pow3 b}-I)$
$$V_{pow3(a)}=\left( \begin{array}{ccc}
0.33& 5.45 & 5.45\\
4.45& 0.80&16.43\\
4.45& 15.43&1.25
\end{array} \right)\ \ \mbox{and}\ \
V_{pow 3(b)}=\left( \begin{array}{ccc}
1.25& 7.00 & 7.00\\
6.00 & 0.80&16.43\\
6.00& 15.43&0.33
\end{array} \right)$$ and then $$tr\left(ASCOV(\sqrt{n} \
\mbox{vec}(\off(\hat{\Gamma}_{pow3(a)})))\right)
%=(4.45 + 4.45 + 5.45
%+ 15.43 + 5.45 + 16.43)
=51.66,$$ and $$tr\left(ASCOV(\sqrt{n} \
\mbox{vec}(\off(\hat{\Gamma}_{pow3(b)})))\right)
%=(6.00 + 6.00 + 7.00
%+ 15.43 + 7.00 + 16.43)
=57.86.$$ Finally,
%The asymptotic variance matrices of
%$\sqrt{n}(\hat{\Gamma}_{tanh a}-I)$ and $\sqrt{n}(\hat{\Gamma}_{tanh
%b}-I)$
$$V_{tanh(a)}=\left( \begin{array}{ccc}
0.33& 7.75 & 7.75\\
6.75 & 0.80&11.37\\
6.75& 10.37 & 1.25
\end{array} \right),
V_{tanh(b)}=\left( \begin{array}{ccc}
1.25& 3.01 & 3.01\\
2.01 & 0.80&11.37\\
2.01& 10.37&0.33
\end{array} \right)$$
which gives $$tr\left(ASCOV(\sqrt{n} \
\mbox{vec}(\off(\hat{\Gamma}_{tanh(a)})))\right)
%=6.75 + 6.75 + 7.75
%+ 10.37 + 7.75 + 11.37
=50.74$$ and $$tr\left(ASCOV(\sqrt{n} \
\mbox{vec}(\off(\hat{\Gamma}_{tanh(b)})))\right)
%=2.01 + 2.01 + 3.01
%+ 10.37 + 3.01 + 11.37
=31.78.$$ There are quite big differences in the asymptotic behavior
of the fastICA estimates only depending on the order in which the
sources are found.  Note also that the variances of the diagonal
elements of $\hat\Gamma$  are equal for all the estimates studied
here. They are simply the limiting variances of the sample variances
of the standardized independent components divided by $4$ as, in all
the cases, the regular covariance matrix is used to whiten the data.
The variances of the diagonal elements of $\hat\Gamma$ are then not
used in the comparison.

\begin{figure}%[h!]
\begin{center}
\includegraphics[angle=270,width=0.6\textwidth]{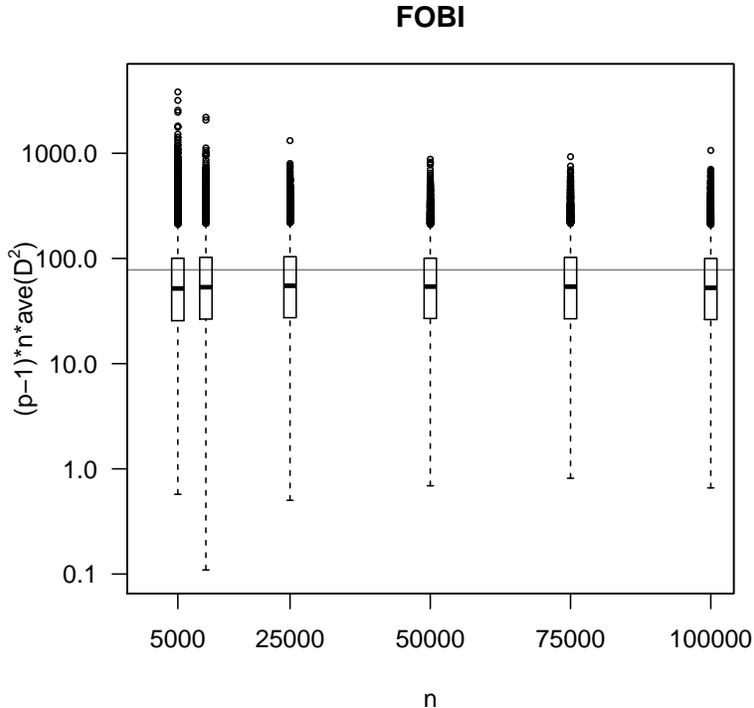}
\caption{Boxplots for $n(p-1)\hat D^2$ based on the FOBI estimate
for different sample sizes $n$  and 10000 repetitions  on log scale. The three
independent components have  Laplace, logistic and $beta(3,3)$
distributions.  The horizontal  line gives the limiting mean value.} \label{boxplots1}
\end{center}
\end{figure}

\begin{figure}%[h!]
\begin{center}
\includegraphics[angle=270,width=0.6\textwidth]{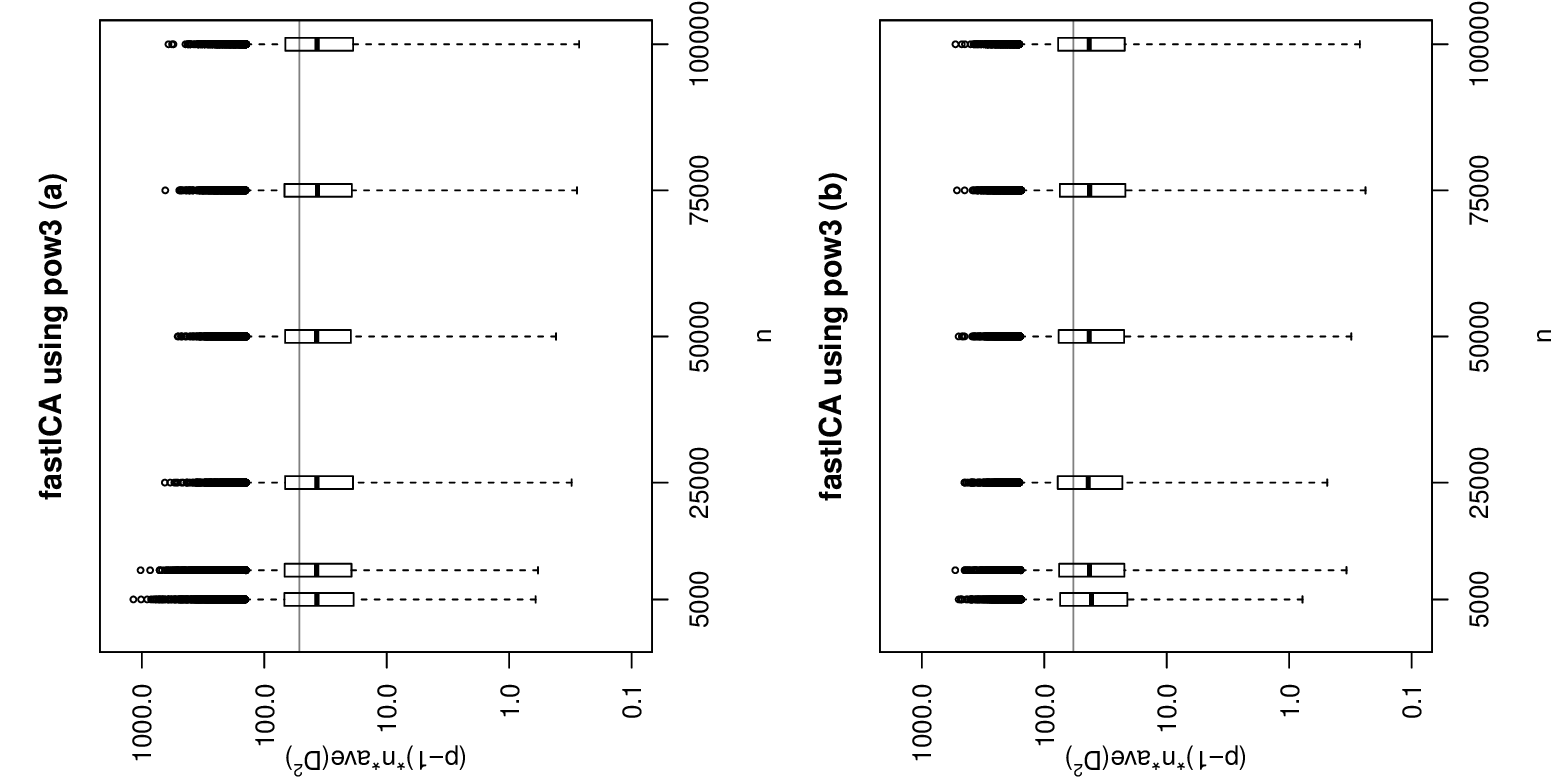}
\caption{ Boxplots for $n(p-1)\hat D^2$ based on the fastICA
estimates pow3(a) and pow3(b) for different sample sizes $n$  and
10000 repetitions on log scale. The three independent components have  Laplace,
logistic and $beta(3,3)$ distributions.  The horizontal  line gives
the limiting mean value. } \label{boxplots2}
\end{center}
\end{figure}

\begin{figure}%[h!]
\begin{center}
\includegraphics[angle=270,width=0.6\textwidth]{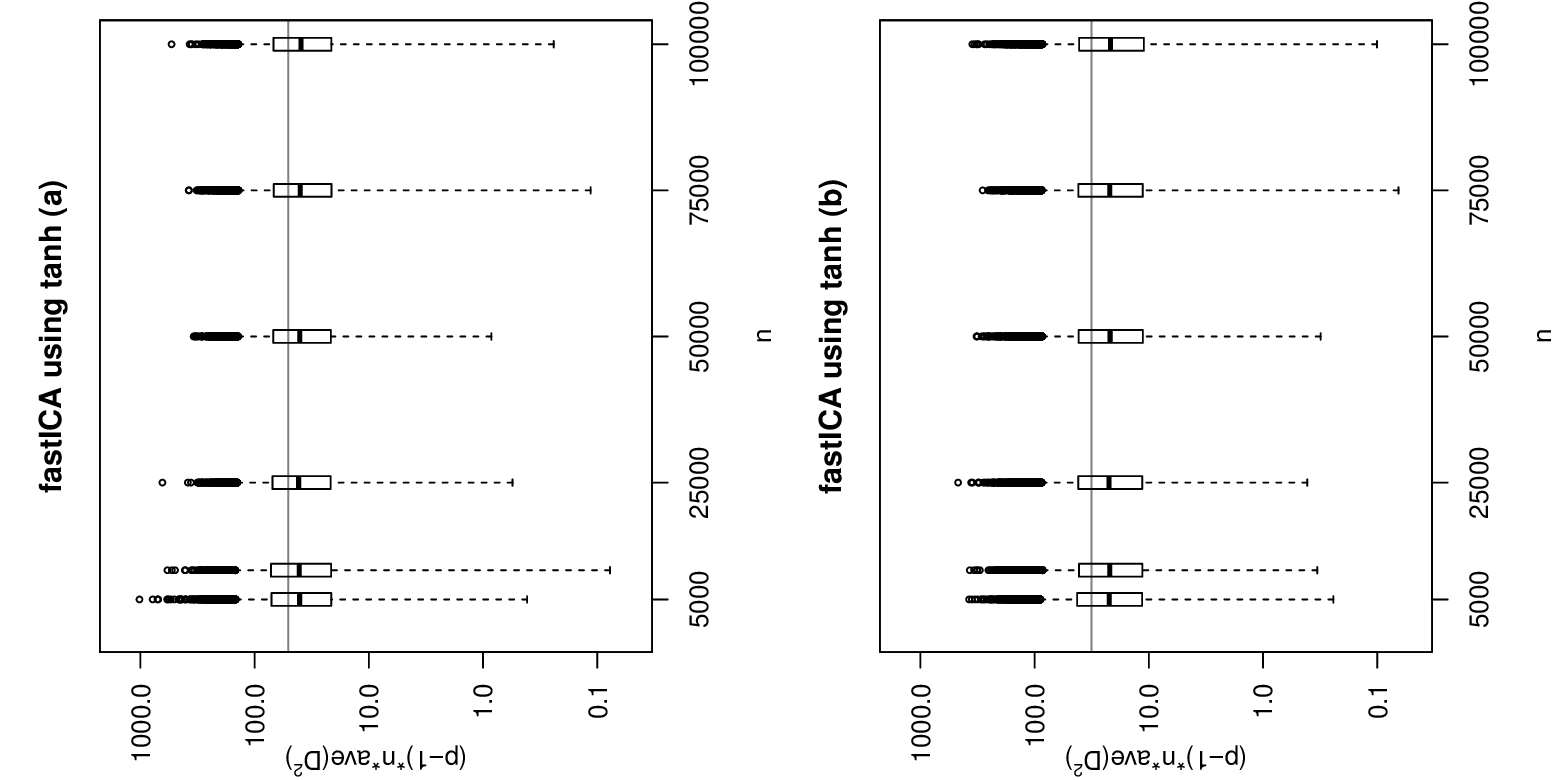}
\caption{ Boxplots for $n(p-1)\hat D^2$ based on the fastICA
estimates tanh(a) and tanh(b) for different sample sizes $n$  and
10000 repetitions on log scale. The three independent components have  Laplace,
logistic and $beta(3,3)$ distributions.  The horizontal line gives
the limiting mean value. } \label{boxplots3}
\end{center}
\end{figure}

\begin{figure}%[h!]
\begin{center}
\includegraphics[angle=270,width=0.55\textwidth]{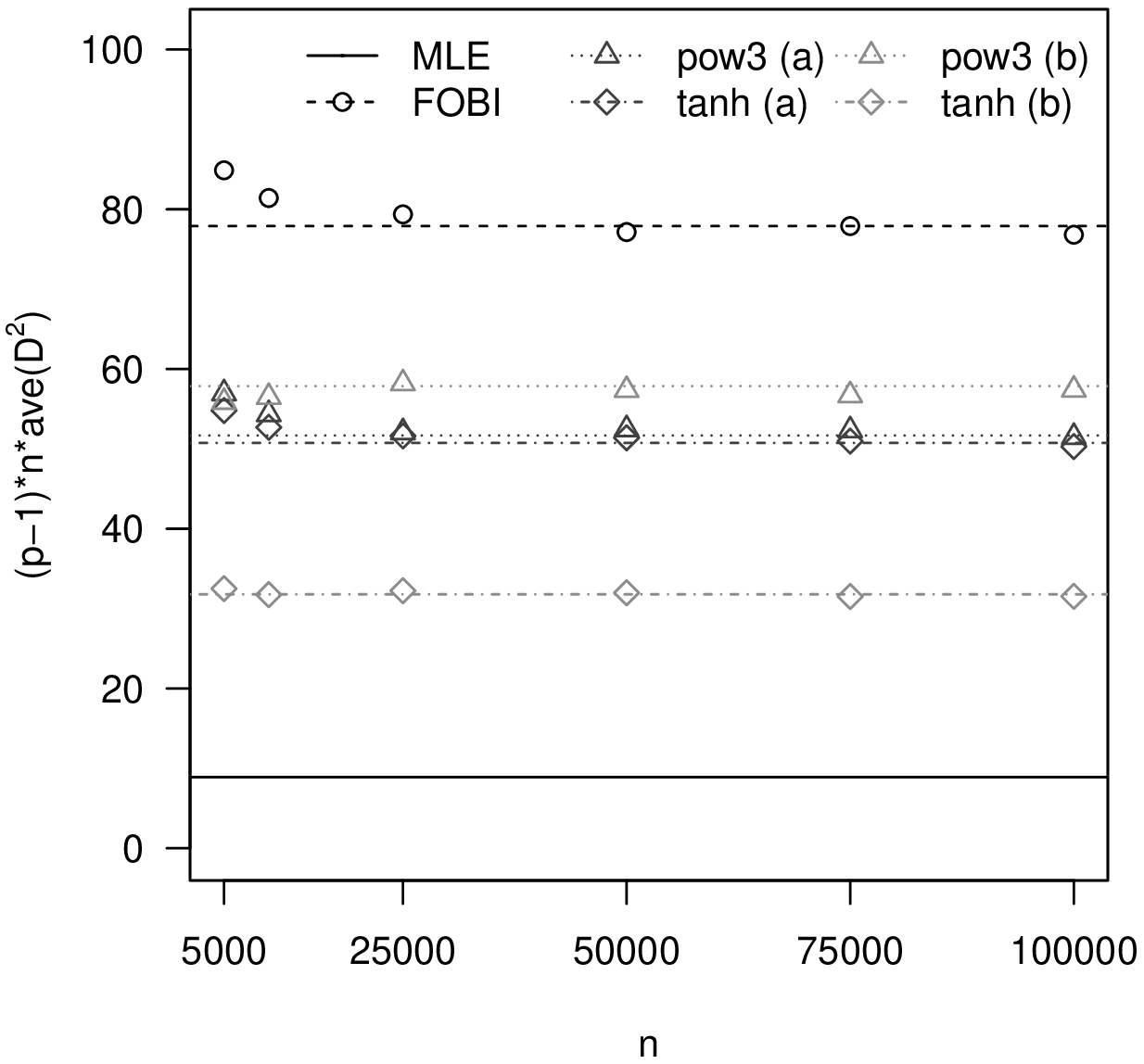}
\caption{The estimated mean values of $n(p-1) \hat D^2$  for the
estimates FOBI, pow3(a), pow3(b), tanh(a) and tanh(b). The dashed
horizontal lines give the corresponding limiting mean values.
 The solid horizontal  line is the limiting mean for the MLE (with known marginal distributions).} \label{boxplotsall}
\end{center}
\end{figure}

Boxplots in Figures \ref{boxplots1}, \ref{boxplots2}  and
\ref{boxplots3} illustrate the finite-sample behavior of the index
for different estimates. The horizontal lines give the limiting
mean values on a log scale. The FOBI estimate is known to
converge in distribution to a multivariate normal distribution, but
the convergence is very slow. The distributional convergence of
$n(p-1)\hat D^2$ is then also slow as is seen from Figure
\ref{boxplots1}. What is interesting, is that the speed of
convergence  of the distribution (not only the covariance structure)
of the fastICA estimate seems to depend on the order of the found
sources,  see Figure \ref{boxplots2} and Figure \ref{boxplots3}. The
distributional convergence of  $n(p-1)\hat D^2$ for tanh(b) seems to
be  faster than that for  tanh(a), see Figure \ref{boxplots3}. The
same is true for pow3(b) and pow3(a) as well, see Figure
\ref{boxplots2}. The estimated means of $n(p-1)\hat D^2$ for
different estimates $\hat\Gamma$ are compared in Figure
\ref{boxplotsall} again with asymptotic horizontal lines. The
performance of the FOBI estimate is clearly worst. The MLE with the
assumption that the marginal distributions are known provides the
Cramer-Rao lower bound  for the limiting mean,
see \cite{OllilaKimKoivunen:2008}. The order in which the sources
are found seems to have a huge effect on the performance of the
fastICA estimate. If the sources are found in the order $beta(3,3)$,
logistic, and Laplace, there is no big difference between choices
pow3: $g(z)=z^3$ and tanh: $g(z)=tanh(z)$. If the order is Laplace,
logistic, and $beta(3,3),$ the  estimate tanh perform very well
while the estimate pow3 gets worse.

\section{Summary}

Independent component analysis (ICA) has gained increasing interest in various fields of applications in recent years.
As far as we know, this paper provides the first rigorous (mathematical) definition of the IC functional. The functional
is defined in a general semiparametric IC model and is independent from the parametrization of the model. %Some popular ICA procedures
%such as JADE are not affine equivariant, and therefore not members in our family of IC functionals.

The deflation-based FastICA algorithm is  one of the most popular ICA algorithms. Several superficial attempts to find the  limiting distribution and  limiting covariance
matrix of the FastICA mixing matrix estimate can be found in the literature (see e.g. \cite{TichavskyKoldovskyOja:2005}, \cite{ShimizuEtAl:2006}, \cite{ReyhaniYlipaavalniemiVigarioOja:2012}). The correct  limiting covariance matrix was found however quite recently in \cite{Ollila:2010}. In this paper we provide the assumptions needed for the limiting multivariate  normality.

For several popular ICA procedures, the statistical  properties are still unknown, and their performances are compared using different performance criteria in simulation studies. In this paper we discuss several criteria in detail and suggest a new performance index with an easy interpretation.
The asymptotic behavior of the new index  depends in a natural way on the eigenvalues of the limiting covariance  matrix of an unmixing matrix estimate.
This is illustrated in a small simulation study with some deflation-based FastICA estimates and with FOBI estimate. We did not use other ICA procedures in our study
as, for other estimates proposed in the literature, the limiting properties are still unknown and/or their implementations cannot deal with the sample sizes of our study.
Note also that the new index can also be computed using the correlation matrix between the estimated and true sources. In that case the index has a nice connection
to the mean-squared error as discussed in \cite{NordhausenOllilaOja:2010}.

 The theory presented in this paper has also important practical implications.  For example, \cite{NordhausenIlmonenMandalOjaOllila:2011} introduces
a new reloaded deflation-based FastICA algorithm that, using a preliminary estimate and the results here,  extracts the sources in an optimal order to minimize the
trace of the limiting covariance matrix.

\section*{Appendix}
\subsection*{The Proof of Theorem \ref{main-canonical}}
Assume that
$\sqrt{n} \ \mathrm{vec} (\hat G-I_p)\to_d N_{p^2}(0,
\Sigma_\Gamma) $.  Now it follows directly from  \cite[Theorem 3.1]{tan} that the limiting distribution of
$
n\cdot D^2(\hat G)
$
 is that of
$\sum_{i=1}^k \delta_i\chi^2_i$ where
$\chi^2_1,....,\chi^2_k$ are independent chi squared variables with
one degree of freedom, and $\delta_1,...,\delta_k$ are the $k$
nonzero eigenvalues (including all algebraic multiplicities) of $\Sigma_\Gamma$.

\subsection*{The Proof of Theorem \ref{AdjEst}}
For simplicity, we consider the elements $\hat\Gamma^*_{11}$ and  $\hat\Gamma^*_{12}$ only. The proofs for other elements
are similar.  First note that
\begin{eqnarray*}
\sqrt{n}\left( (\hat \Gamma\hat S\hat  \Gamma')_{11}-D_{11}^{-2} \right)
&=& \sqrt{n}\left( \sum_{i=1}^p\sum_{j=1}^p \hat \Gamma_{1i} \hat S_{ij} \hat  \Gamma_{1j}-D_{11}^{-2}\right)\\
&=& \sqrt{n}\left( 2 (\hat \Gamma_{11}-D_{11}^{-1})D_{11}^{-1}+(\hat S_{11}-1)D_{11}^{-2}\right)+ o_P(1).
\end{eqnarray*}
But then
\[
\sqrt{n}\left((\hat \Gamma\hat S\hat  \Gamma')_{11}^{-1/2}-D_{11} \right)=
-\sqrt{n}\left( D_{11}^2(\hat \Gamma_{11}-D_{11}^{-1})+\frac {D_{11}}2 (\hat S_{11}-1) \right)+o_P(1)
\]
and
\begin{eqnarray*}\sqrt{n}(\hat \Gamma_{11}^*-1) &=&
\sqrt{n}\left((\hat \Gamma\hat S\hat  \Gamma')_{11}^{-1/2}\hat \Gamma_{11}-1 \right) \\
&=&
\sqrt{n}\left( ((\hat \Gamma\hat S\hat  \Gamma')_{11}^{-1/2}-D_{11})D_{11}^{-1}+D_{11}(\hat \Gamma_{11}-D_{11}^{-1})  \right)
\\ &\ &  +o_P(1)\\
&=&-\frac {\sqrt{n}}2 (\hat S_{11}-1)+o_P(1).
\end{eqnarray*}
Finally,
\[
\sqrt{n} \hat \Gamma_{12}^* =
\sqrt{n}\left((\hat \Gamma\hat S\hat  \Gamma')_{11}^{-1/2}\hat \Gamma_{12} \right)
=
\sqrt{n}D_{11}^{-1}\hat \Gamma_{12}+o_P(1).
\]

\subsection*{The Proof of Theorem \ref{main-AdjEst}}
Assume that
$\sqrt{n} \ \mathrm{vec} (\hat G-I_p)\to_d N_{p^2}(0,
\Sigma_{\Gamma^*}) $, and let  $D_{p,p}=\sum_{i=1}^p (e_i e_i^{T})\otimes (e_i
e_i^{T})$. Now $ \mathrm{vec} (\mbox{off}(\hat G))= (I_{p^2}-D_{p,p})\mathrm{vec} (\hat G-I_p)$ and thus $\sqrt{n} \ \mathrm{vec} (\mbox{off}(\hat G))\to_d N_{p^2}(0,
(I_{p^2}-D_{p,p})
\Sigma_{\Gamma^*} (I_{p^2}-D_{p,p})). $ Now it follows from  \cite[Theorem 3.1]{tan} that the limiting distribution of
$
n ||\mbox{off}(\hat G)||^2
$
 is that of
$\sum_{i=1}^k \delta_i\chi^2_i$ where
$\chi^2_1,....,\chi^2_k$ are independent chi squared variables with
one degree of freedom, and $\delta_1,...,\delta_k$ are the $k$
nonzero eigenvalues (including all algebraic multiplicities) of
\[
(I_{p^2}-D_{p,p})
\Sigma_{\Gamma^*} (I_{p^2}-D_{p,p}).
\]

\subsection*{The Proof of Theorem \ref{stand2_Th}}
The proof follows from the fact that $\sqrt{n}(\hat D_{ii}\hat G_{ii}-1)=o_P(1)$. Then
also $\sqrt{n}\hat D_{ii}\hat G_{ij}= \sqrt{n} D_{ii}\hat G_{ij}+ o_P(1)$ for all $i\ne j$.

\subsection*{The Proof of Lemma \ref{trace}}
Let $A=(a_{ij})$ be a $p\times p$ matrix having at least one nonzero element in each row and let $\tilde{A}=(\tilde{a}_{ij})=\frac{a_{ij}^2}{\sum_{k=1}^pa_{ik}^2}.$ Let $\mathcal L$ denote the set of all nonsingular $p\times p$ diagonal matrices and let $L=(l_{ij}) \in \mathcal L.$  Now
$$\| LA-I_p\|^2=\sum_{i=1}^p\sum_{j=1}^pl_{ii}^2 a_{ij}^2 -2\sum_{i=1}^pl_{ii} a_{ii} + p$$ and $$\frac{\partial}{\partial l_{ii}}\| LA- I_p\|^2=2(l_{ii}\sum_{j=1}^pa_{ij}^2-a_{ii}).$$
The derivatives are zero with choices $$l_{ii}=\frac{a_{ii}}{\sum_{j=1}^pa_{ij}^2}$$ and the value of $\| L A-I_p\|^2$ is then $$p-\sum_{i=1}^p\frac{a_{ii}^2}{\sum_{j=1}^pa_{ij}^2}.$$ Let $\mathcal P$ denote the set of all $p\times p$ permutation matrices. Now it follows that
if $\hat{G} =
\hat{\Gamma}\Omega$, and $\tilde{ G}_{ij}={\hat
G_{ij}^2}/{\sum_{k=1}^p \hat G_{ik}^2}$, $i,j=1,...,p$, then the
minimum distance index can be written as
$$
\hat D= D(\hat{G}) = \frac 1{\sqrt{p-1}} \left(p - \max_{
P\in\mathcal P}\left( \mbox{tr}(P \tilde{G})\right) \right)^{1/2}.$$

\subsection*{The Proof of Theorem \ref{fourcond}}
Let $A=(a_{ij})$ be a $p\times p$ matrix having at least one nonzero element in each row. Let $\tilde{A}=(\tilde{a}_{ij})$ with $\tilde{a}_{ij}=\frac{a_{ij}^2}{\sum_{k=1}^pa_{ik}^2}.$
Let $\mathcal P$ denote the set of all $p\times p$ permutation matrices.
Now the shortest squared distance $D^2(A)=\frac 1{p-1}(p-\max_{
P\in\mathcal P}( \mbox{tr}(P\tilde{A}))).$ (See the proof of Lemma \ref{trace}.) Consider now $\mbox{tr}(P\tilde{A}),$ where $\tilde{a}_{ij}\geq0,$ for all $i,j$ and $\sum_{j=1}^p\tilde{a}_{ij}=1.$ Now clearly the maximum value of $\max_{
P\in\mathcal P}\mbox{tr}(P\tilde{A})$ is $p$ and it is attained if and only if $\tilde{A}$ is a permutation matrix. Since $\tilde{A}$
is a permutation matrix if and only if $A \sim I_p,$
we have now proven that
$D^2(A)\ge 0$ for all $A$ and that
$D^2(A)=0$ if and only if $A \sim I_p$.

\bigskip

For the minimum value of $\max_{
P\in\mathcal P}\mbox{tr}(P\tilde{A})$  note that
$$\max_{
P\in\mathcal P}\mbox{tr}(P\tilde{A}) \geq \frac{1}{p!}\sum_{P\in\mathcal P}\mbox{tr}(P\tilde{A})=\frac{1}{p}\sum_{i=1}^p\sum_{j=1}^p\tilde{a}_{ij}=1.$$
If $\max_{
P\in\mathcal P}\mbox{tr}(P\tilde{A})=1,$ then $\mbox{tr}(P\tilde{A})=1$ for all permutation matrices $P.$
Since all row sums of $\tilde{A}$ are one, if rows $i_1\neq i_2$ of $\tilde{A}$ are different,
there have to exist indices $j_1\neq j_2$ such that $\tilde{a}_{i_1j_1} > \tilde{a}_{i_2j_1},$ and that
$\tilde{a}_{i_1j_2} < \tilde{a}_{i_2j_2}.$ Let now $P_1$ and $P_2$ denote permutation matrices which are identical in all rows
$i\not \in \{i_1, i_2\}$ and in all columns $j\not \in \{j_1, j_2\},$ and let the elements $i_1j_1$ and $i_2j_2$ of $P_1$ be equal to one, and let
the elements $i_1j_2$ and $i_2j_1$ of $P_2$ be equal to one.
Then $\mbox{tr}(P_1\tilde{A})>\mbox{tr}(P_2\tilde{A})$ contradicting the fact that $\mbox{tr}(P\tilde{A})$ is identical for all permutation matrices $P.$
Hence $A \sim 1_p a^T$ for
some $p$-vector $a$.
%But then
%$$\tilde{a}_{ii}+\tilde{a}_{jj}=\tilde{a}_{ij}+\tilde{a}_{ji},$$ for all $i,j$ and summing over $j$ gives
%$$\tilde{a}_{ii}=\frac{1}{p}\sum \tilde{a}_{ji}.$$ As this holds for any permuted versions of $\tilde{A}$ as well,
%all the row vectors of $\tilde{A}$ must be identical.
We have now proven that
$1\ge D^2(A)$ for all $A$ and that
$D^2(A)=1$ if and only if $A \sim 1_p a^T$ for
some $p$-vector $a$.

\bigskip

Assume now that $a_{ij}\leq 1, i\neq j$ and let $B =I_p + c \ \off(A),$ where $c\in [0,1]$ and let $\tilde{B}=(\tilde{b}_{ij})=\frac{b_{ij}^2}{\sum_{k=1}^pb_{ik}^2}$. Then $\max_{
P}\mbox{tr}(P\tilde{B})=\sum_{i=1}^p\frac{1}{c^2\sum_{j=1, j\neq i}^pa_{ij}^2 + 1}.$ Now clearly $\max_{
P}\mbox{tr}(P\tilde{B})$ decreases when $c$ increases. This proves that the function $c\to D^2(I_p + c \ \off(A))$ is increasing in $c\in [0,1]$ for all matrices $A$ such that
$A_{ij}^2 \leq 1$, $i\ne j$.
\subsection*{The Proof of Theorem \ref{main}}
Let $\Gamma(F_{
x})=\Omega=C(F_{x})= I_p$ and let $
\sqrt{n} \ \mathrm{vec} (\hat{\Gamma}-I_p)\to_d N_{p^2}(0,
\Sigma) $.
Let $\mathcal P$ denote the set of all $p\times p$ permutation matrices and let $\mathcal L$ denote the set of all nonsingular $p\times p$ diagonal matrices.

\bigskip

We have $$\hat D=\frac 1{\sqrt{p-1}} \inf_{C\in \mathcal C} \| C
\hat{\Gamma} -I_p\|=\frac 1{\sqrt{p-1}} \min_{P\in \mathcal P}( \inf_{L\in \mathcal L} \| LP
\hat{\Gamma} -I_p\|).$$ Let $$\hat{P}_m=\argmin_{P\in \mathcal P}(\inf_{L\in \mathcal L} \| LP
\hat{\Gamma} -I_p\|), \ \hat{L}_m=\arginf_{ L\in \mathcal L} \|  L\hat{ P}_m
\hat{\Gamma} -I_p\|,$$ $$P_m=\argmin_{P\in \mathcal P}(\inf_{L\in \mathcal L} \| LP
\Gamma -I_p\|), \ L_m=\arginf_{L\in \mathcal L} \| LP_m
\Gamma -I_p\|$$ and for all $P\in \mathcal P$ let  $\hat{L}_P=\arginf_{L\in \mathcal L} \| LP
\hat{\Gamma} -I_p\|$ and $L_P=\arginf_{L\in \mathcal L} \| LP
\Gamma -I_p\|.$  Now for all $P\in \mathcal P,$ $(\hat{L}_P)_{ii}=\hat{B}_{ii}/\sum_{j=1}^p \hat{B}_{ij}^2,$ where $\hat{B}_{ij}=(\hat{ B}_P)_{ij}=(P\hat{\Gamma})_{ij}$ and
$(L_P)_{ii}=B_{ii}/\sum_{j=1}^p B_{ij}^2,$ where $B_{ij}=(B_P)_{ij}=(P\Gamma)_{ij},$
(see the proof of Lemma \ref{trace}). Let $P\in \mathcal P.$ Since $(\hat{\Gamma} - \Gamma)\stackrel{P}{\rightarrow} \bar{0},$ it now follows from the continuous mapping theorem that also $(P\hat{\Gamma} - P\Gamma)\stackrel{P}{\rightarrow} \bar{0}$ and thus $(\hat{L}_P - L_P)\stackrel{P}{\rightarrow} \bar{0}.$ Since $(\hat{L}_P - L_P)\stackrel{P}{\rightarrow} \bar{0}$ holds for all $P\in \mathcal P,$ it follows that $(\hat{P}_m - P_m)\stackrel{P}{\rightarrow} \bar{0}$ and
 $(\hat{L}_m - L_m)\stackrel{P}{\rightarrow} \bar{0}$  as well.
\bigskip
Clearly $$P_m=L_m=I_p.$$ Since $P_m$ and $\hat{P}_m$ are discrete, we now have, by using Slutsky's theorem, that  $$\sqrt{n} \ \mathrm{vec} (\hat{L}_m \hat{P}_m \hat{ \Gamma}-I_p)=\sqrt{n} \ \mathrm{vec} (\hat{L}_m-I_p)+\sqrt{n} \ \mathrm{vec} (\hat{\Gamma}-I_p) + o_P(1).$$

\bigskip

Consider now $$\sqrt{n} \ \mathrm{vec} (\hat{L}_m-I_p).$$ Let $\hat{A}=(\hat a_{ij})=\hat{P}_m\hat{\Gamma}$ and
define diagonal matrices $\hat{D}_a =(\hat{D}_a)_{ii}=\hat a_{ii},$ $\hat{D}_b =(\hat{D}_b)_{ii}= \frac{1}{\sum_{j=1}^p\hat a_{ij}^2}.$
 Now $$\sqrt{n}(\hat{L}_m-I_p)= \sqrt{n}(\hat{D}_a - I_p)\hat{D}_b + \sqrt{n}(\hat{D}_b-I_p)$$ and it follows from the convergency of $\hat{P}_m$ and $\hat{\Gamma}$ that $(\hat{D}_a - I_p)\stackrel{P}{\rightarrow} \bar{0}$  and $(\hat{D}_b - I_p)\stackrel{P}{\rightarrow} \bar{0}.$

\bigskip

Consider now the $ii$ element of the matrix $\sqrt{n}(\hat{D}_b-I_p).$ We have $\sqrt{n}( \frac{1}{ \sum_{j=1}^p\hat a_{ij}^2}-1)=\sqrt{n}\frac{1-\sum_{j=1}^p\hat a_{ij}^2}{\sum_{j=1}^p\hat a_{ij}^2}.$ It now follows from our assumptions and discreteness of  $\hat{P}_m$ and $P_m$ that each $\sqrt{n}\hat a_{ij}, \ i\neq j$ converges to normal distribution with zero mean. Now each  $n\hat a_{ij}^2,\ i\neq j$ converges in distribution to a $\chi^2$ variable and thus each $\sqrt{n}\hat a_{ij}^2, \ i\neq j$ converges in probability to zero and $(\sqrt{n}\frac{1-\sum_{j=1}^p\hat a_{ij}^2}{\sum_{j=1}^p\hat a_{ij}^2}-\sqrt{n}\frac{1-\hat a_{ii}^2}{\sum_{j=1}^p\hat a_{ij}^2})$ $=(\sqrt{n}\frac{1-\sum_{j=1}^p\hat a_{ij}^2}{\sum_{j=1}^p\hat a_{ij}^2}-(\frac{1+\hat a_{ii}}{-\sum_{j=1}^p\hat a_{ij}^2}(\sqrt{n}(\hat a_{ii}- 1))))$ converges in probability to zero as well. Now since $\frac{1+\hat a_{ii}}{-\sum_{j=1}^p\hat a_{ij}^2}=-2+ o_P(1),$ it follows from Slutsky's theorem that $$\sqrt{n} \ \mathrm{vec} (\hat{L}_m-I_p) = -\sqrt{n}\ \mathrm{vec}(\hat{D}_a - I_p)+ o_P(1).$$ Since $(\hat{\Gamma}-I_p)= (\hat{P}_m\hat{\Gamma}-I_p) + ((\hat{P}_m-I_p)\hat{\Gamma})$, we now have by Slutsky's theorem and discreteness of $\hat{P}_m$ that $$\sqrt{n}\mathrm{diag}(\hat{\Gamma}-I_p)=\sqrt{n}\mathrm{diag}(\hat{P}_m\hat{\Gamma}-I_p)+ o_P(1).$$ Since $$\sqrt{n}(\hat{D}_a - I_p)=\sqrt{n}\mathrm{diag}(\hat{P}_m\hat{\Gamma}-I_p),$$ we conclude, using Slutsky's theorem again, that
$$\sqrt{n} \ \mathrm{vec} (\hat{L}_m \hat{P}_m \hat{\Gamma}-I_p)\stackrel{d}{\rightarrow} N(\bar{0}, \Sigma_2),$$ where
\[\Sigma_2=ASCOV(\sqrt{n} \
\mbox{vec}(\off(\hat{\mathrm\Gamma})))=(I_{p^2}-D_{p,p})
\Sigma (I_{p^2}-D_{p,p})
\]
with $D_{p,p}=\sum_{i=1}^p (e_i e_i^{T})\otimes (e_i
e_i^{T})$.
Thus
$$n\hat D^2=\frac n{p-1} \|  \off(\hat{\Gamma})\|^2 + o_P(1)$$ and it follows from \cite[Theorem 3.1]{tan}, that
the limiting distribution of $n\hat D^2$ is that of
$(p-1)^{-1}\sum_{i=1}^k \delta_i\chi^2_i$ where
$\chi^2_1,....,\chi^2_k$ are independent chi squared variables with
one degree of freedom, and $\delta_1,...,\delta_k$ are the $k$
nonzero eigenvalues (including all algebraic multiplicities) of $\Sigma_2.$

%\section*{Acknowledgements}
%We would like to thank the Associate Editor and the anonymous referee. Their comments and suggestions led to a considerable improvement of the present paper.

%\section*{Acknowledgements}
% This research was supported by the Academy of Finland.

%%%%%%%%%%%%%%%%%%%%%%%%%%%%%%%%%%%%%%%%%%%%%%%%%%%%%%%%%%%%%%%%%%%%%%%%%%%%%%%%%%%%%%%%%%%%%%%%%%%%
%%%%%%%%%%%%%%%%%%%%%%%%%%%%%%%%%%%%%%%%%%%%%%%%%%%%%%%%%%%%%%%%%%%%%%%%%%%%%%%%%%%%%%%%%%%%%%%%%
%%%%%%%%%%%%%%%%%%%%%%%%%%%%%%%%%%%%%%%%%%%%%%%%%%%%%%%%%%%%%%%%%%%%%%%%%%%%%%%%%%%%%%%%%%%%%%%%%%%%%%%%%%%%%

%%%%%%%%%%%%%%%%%%%%%%%%%%%%%%%%%%%%%%%%%%%%%%%%%%%%%%%%%%%%%%%%%%%%%%%%%%%%%%%%%%%%%%%%%%%%%%%%%%%%
%%%%%%%%%%%%%%%%%%%%%%%%%%%%%%%%%%%%%%%%%%%%%%%%%%%%%%%%%%%%%%%%%%%%%%%%%%%%%%%%%%%%%%%%%%%%%%%%%
%%%%%%%%%%%%%%%%%%%%%%%%%%%%%%%%%%%%%%%%%%%%%%%%%%%%%%%%%%%%%%%%%%%%%%%%%%%%%%%%%%%%%%%%%%%%%%%%%%%%%%%%%%%%%

%############################################################################

\end{document}